\documentclass[fleqn,usenatbib]{mnras}

\usepackage{mathptmx}
\usepackage{txfonts}
\usepackage[T1]{fontenc}
\usepackage{ae,aecompl}
\usepackage{multirow}

\usepackage{graphicx}	
\newcommand{\bz}{$\langle B_z \rangle$}

\title[Coherent radio emission from HD\,35298]{The fifth main sequence magnetic B-type star showing coherent radio emission: is this really a rare phenomenon?}

\author[B. Das et al.]{
Barnali Das,$^{1}$\thanks{E-mail: barnali@ncra.tifr.res.in}
Poonam Chandra,$^{1}$
Matt E. Shultz,$^{2}$
and Gregg A. Wade$^{3}$
\\
$^{1}$National Centre for Radio Astrophysics, Tata Institute of Fundamental Research,  Pune University Campus, Pune-411007, India\\
$^{2}$Annie Jump Cannon fellow, Department of Physics and Astronomy, University of Delaware, 217 Sharp Lab, Newark, DE 19716, USA\\
$^{3}$Department of Physics and Space Science, Royal Military College of Canada, PO Box 17000, Station Forces, Kingston, ON K7K 7B4, Canada
}

\date{Accepted XXX. Received YYY; in original form ZZZ}

\pubyear{2019}
\hypersetup{draft} 
\begin{document}
\label{firstpage}
\pagerange{\pageref{firstpage}--\pageref{lastpage}}
\maketitle

\begin{abstract}
We report the discovery of intense, highly directional radio emission from the Bp star HD\,35298, 
which we interpret as the consequence of Electron Cyclotron Maser Emission (ECME). The star was observed with the Giant Metrewave Radio Telescope near the rotational phases of both magnetic nulls in band 4 ($550-750$ MHz) and one of the nulls in band 5 ($1060-1460$ MHz).
In band 4, we observed flux density enhancement in both circular polarizations near both magnetic nulls.
The sequences of arrival of the left and right circularly polarized pulses are opposite near the two nulls.
In band 5, we did not have circular polarization information and hence measured only the total intensity lightcurve, which also shows enhancement around the magnetic null. The observed sequence of the circular polarization signs in band 4, compared with the longitudinal magnetic field curve, is able to locate the hemisphere from which ECME arises. This observational evidence supports the scenario of ECME in the ordinary mode, arising in a magnetosphere shaped like an oblique dipole.
HD\,35298 is the most slowly rotating and most distant main sequence magnetic star from which ECME has been observed.


\end{abstract}




\section{Introduction}
Hot magnetic stars are expected to exhibit radio emission primarily by the gyrosynchrotron mechanism. Such emission arises due to the interaction of a radiatively driven stellar wind with the stellar magnetic field \citep[e.g.][]{andre88,trigilio04}. Among these stars, a very small number have been found to produce coherent radio emission in their magnetospheres. The first such star discovered was CU\,Vir \citep{trigilio00}. The authors observed highly circularly polarised ($\sim$100\%) pulses from this star at 1.4 GHz, which was absent at 5 GHz and above \citep{leto06}. The rotational phases corresponding to the arrival of the pulses nearly coincide with the nulls of the star's longitudinal magnetic field (hereafter referred to as magnetic null phases). From these characteristics \citeauthor{trigilio00} proposed the emission mechanism to be Electron Cyclotron Maser Emission (ECME) which is known to produce highly circularly polarized emission. ECME is directed at an angle of $\approx\cos^{-1}(v/c)$ with respect to the magnetic field direction \citep{melrose82}, where $v$ is a characteristic speed of the non-thermal electron distribution responsible for maser amplification and $c$ is the speed of light. For a mildly relativistic electron distribution, this angle is close to $90^\circ$, which explains the occurrence of the pulses near the magnetic null phases. 

The same phenomenon has also been observed from ultracool dwarfs \citep[UCDs, e.g.][]{hallinan06,hallinan07}. The atmospheres of these stars are expected to be neutral and cooler than the higher mass dwarf stars \citep{hallinan07,mohanty03,stelzer06} and the mechanism that accelerates electrons to high energies resulting into auroral radio emission is not yet clear \citep{hallinan15}. This discovery of ECME revealed that UCDs can harbour large-scale magnetic fields as strong as $\sim$ kG \citep{hallinan06,hallinan08}. \cite{hallinan15} also speculated that H$\mathrm{\alpha}$ emission observed from these stars is generated from the same electron beam responsible for auroral radio emission, based on their simultaneous radio and optical observation of the M8.5 dwarf LSR J183513259. This was also supported by the fact that the detection rate of ECME in brown dwarfs increases significantly by biasing the selection of targets in favour of the presence of H$\mathrm{\alpha}$ or optical/IR variability \citep{kao16}.



\begin{table}
{\scriptsize
\caption{Stellar and magnetic properties of HD\,35298  \label{tab:stars}}
\begin{tabular}{lll}
\hline
Physical quantity & Value & Reference\\
\hline
Spectral type & B3 Vw & \cite{ciatti71}\\
Mass $(M/M_\odot)$ & $4.3\pm 0.2$ & Shultz et al. 2019, submitted\\
Radius $(R_*/R_\odot)$ & $2.42\pm 0.06$ & Shultz et al. 2019, submitted\\
Temperature $(T_\mathrm{eff}/\mathrm{kK})$ & $15.8\pm 0.8$ & \cite{shultz19} \\
Rotation period ($P_{\mathrm{rot}}/\mathrm{day}$) & $1.85458(3)$ & \cite{shultz18}\\
Dipole strength $(B_\mathrm{d}/\mathrm{kG})$ & $10.8\pm 0.8$ & Shultz et al. 2019, submitted\\ 
Inclination angle $(i_\mathrm{rot})$ & $64 \pm 3$ & Shultz et al. 2019, submitted\\
Obliquity $(\beta)$ & $78^{+1}_{-4}$ & Shultz et al. 2019, submitted\\
Age ($\log(t/\mathrm{yr})$) &$7.00\pm 0.10$ & \cite{landstreet07}\\ 
\hline
\end{tabular}
}
\end{table}

Currently, only four hot magnetic stars are known to exhibit ECME: CU\,Vir \citep{trigilio00}, HD\,133880 \citep{chandra15,das18}, HD\,142301 \citep{leto19} and HD\,142990 \citep{lenc18,das19}. In this Letter, we report the discovery of the fifth such star: HD\,35298. HD\,35298 is the most slowly rotating hot magnetic star ($P_{\mathrm{rot}}\approx 1.85$ d) in which ECME has so far been detected. 

This Letter is structured as follows: in the next section, we briefly describe HD\,35298 (\S\ref{sec:hd35298}); in the following section (\S\ref{sec:obs}), we describe our observations and data analysis. We then present the results of our analysis  (\S\ref{sec:results}), and finally in \S \ref{sec:disc_conc}, we summarize our conclusions. 

\section{HD 35298}\label{sec:hd35298}
HD\,35298 is a chemically peculiar B type star in the Orion constellation, and was listed as a probable member of the Ori OB 1a cluster by \cite{landstreet07}. In Table \ref{tab:stars}, we listed values of some of the physical parameters for this star. Based on its relatively rapid rotation and strong magnetic field, the star is expected to have a centrifugal magnetosphere \citep{petit13}. Despite this, it does not display H$\mathrm{\alpha}$ emission \citep{petit13}. Its magnetospheric activity was however detected in radio, both in cm \citep{linsky92} and mm \citep{leone04} range. The 6 cm (5 GHz) flux density of this star was reported to be 0.28$\pm$0.06 mJy \citep{linsky92} and that at 3.4 mm (87.7 GHz) was reported to be 1.61$\pm$0.43 mJy \citep{leone04}. However, the corresponding rotational phases were not reported, hence we cannot extract a meaningful spectral index from these data. Moreover, the spectrum is also severely undersampled. Therefore, in this paper we will assume the flux density reported by \cite{linsky92} as the basal (gyrosynchotron) flux density in band 4. This is justified because many magnetic chemically peculiar stars are reported to have flat spectra (spectral index $\approx |0.1|-|0.4|$) over the centimeter wavelength range \citep{drake87,trigilio04,leto06,leto12,leto17,leto18}.

The surface magnetic field of the star is expected to be close to that of a dipole since its longitudinal magnetic field \bz~shows a nearly sinusoidal variation with rotational phase \citep{shultz18}. Although, the \bz variation can be best fitted with a 3rd order sinusoid, the amplitude of the second and third harmonics of \bz~are smaller than that of the fundamental harmonic by more than a factor of ten \citep{shultz18} implying that the deviation of the magnetic field of HD\,35298 from a pure dipole is likely to be small. 

\section{Observations and data reductions}\label{sec:obs}
HD\,35298 was observed on 2018 May 18 and 2018 May 19 at 550--750 MHz (band 4) and on 2019 April 29 at 1060--1460 MHz (band 5) with the upgraded Giant Metrewave Radio Telescope (uGMRT). 
On each day we observed around one of the magnetic null phases which were identified with the help of the \bz~measurements and rotation period reported by \cite{shultz18}. The rotational phase range covered varies between $0.06-0.07$ cycles. The standard flux calibrator 3C48 was used to calibrate the absolute flux density and also as a bandpass calibrator. The phase calibrators used were J0607-085 (band 4), and J0503+020 and J0532+075 (band 5).

The data were analysed using the Common Astronomy Software Applications package \citep[\textsc{casa},][]{mcmullin07}. Dead antennae were identified using the task `plotms' and flagged. The edges of the observing frequency band were also excluded due to very low sensitivity. After that, we ran a \textsc{casa}-based pipeline primarily written for uGMRT data (Ishwara-Chandra et al. in preparation). This pipeline uses standard \textsc{casa} tasks to do flagging, calibration and imaging. The task `flagdata' is used for removing data corrupted by Radio Frequency Interference (RFI). For evaluating the antenna gains, the tasks `gaincal' and `bandpass' are used which, respectively, find the time-dependent and frequency-dependent parts of the antenna gains. The calibrated data are again flagged using `flagdata' and the data are re-calibrated. The calibrated data for the target are separated out and passed to the task `tclean' for imaging. The output of this pipeline is the self-calibrated image of the target-field. 

In addition to running built-in \textsc{casa} tasks to clean the data, we also used an offline flagging routine called `ankflag' (A. Bera \& S. Mondal 2019, in preparation) on the data acquired on 2019 April 29 (band 5). This routine, written primarily for uGMRT data, was used after we found a significant amount of RFI even after running `rflag' (a flagging routine in \textsc{casa}). 

In the self-calibrated image of HD\,35298, we noticed that the field is devoid of any bright source (flux density > 50 mJy). In such a case the assumption that the sky is stationary in time does not hold and consequently the antenna gains obtained from self-calibration cannot be trusted. This situation does not arise in the presence of bright source(s) (whose flux densities are constant over the observation duration) in the field of view. This is because the self-calibration process is primarily driven by the flux density of the bright sources and hence the antenna gains are not affected by the relatively weak time-variability of the target. 

In order to obtain the lightcurves of the target, we therefore self-calibrated each timeslice of length 5 minutes independently. This of course includes the assumption that the flux density variation of the target is not significant over 5 minutes (or over a rotational phase window of width 0.002).


We also observed the star with the legacy GMRT on 2014 November 25 in L-band (1420 MHz) with a bandwith of 33.33 MHz centred around 1387 MHz. We could obtain only the total intensity lightcurve for these data since the GMRT feeds in band 5 are linear and we did not observe polarization calibrator. These data were analysed following a similar procedure.


\begin{figure*}
    \centering
    \includegraphics*[width=0.9\textwidth]{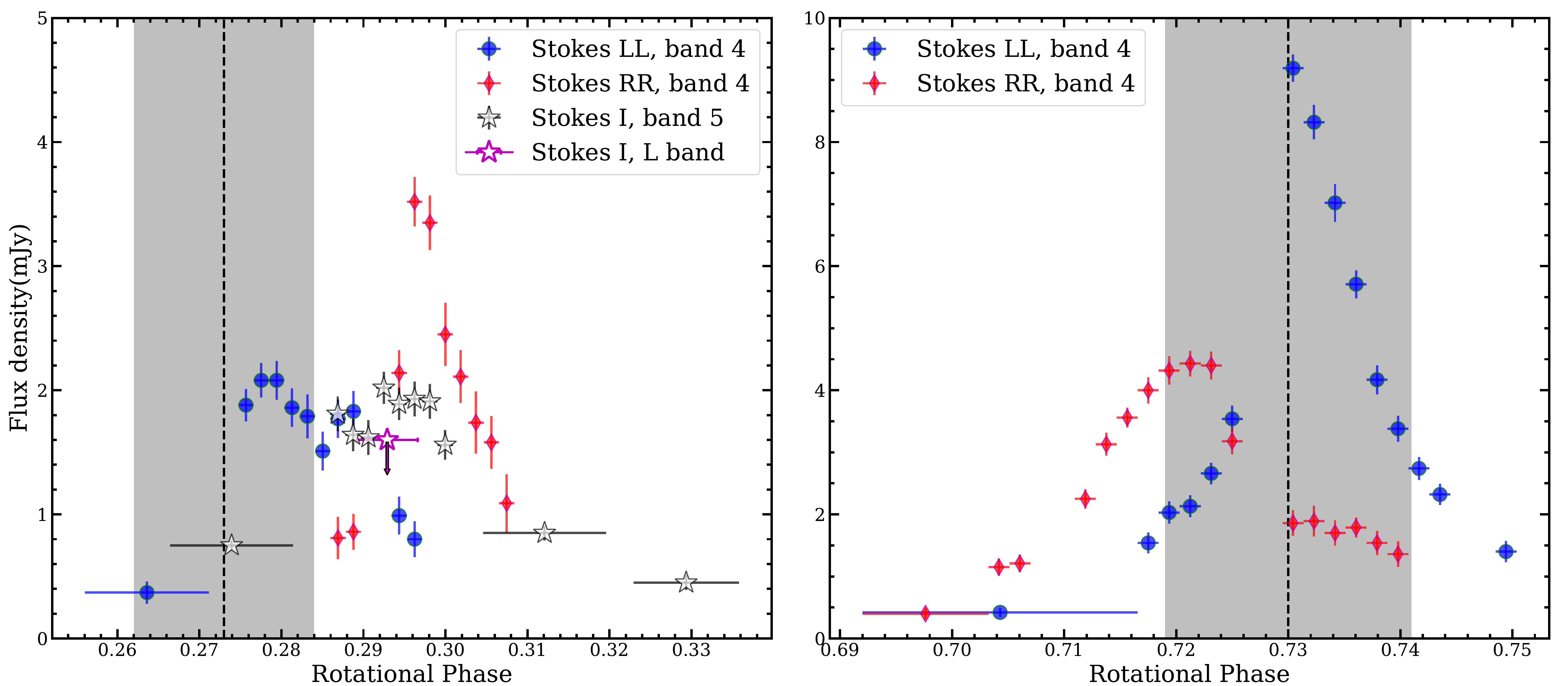}
    \caption{Lightcurve of HD\,35298 near the magnetic null phases 0.273 and 0.730 (right panel). Stokes LL and RR represent LCP and RCP respectively. Stokes I represents total intensity. Band 4 and band 5 data were obtained with the uGMRT in 2018--2019, whereas L-band data were obtained with the legacy GMRT in 2014. The vertical dashed lines represent the location of the magnetic null phases and shaded regions around them represent the associated uncertainties.}
    \label{fig:hd35298}
\end{figure*}

\begin{figure*}
    \centering
    \includegraphics*[width=0.85\textwidth]{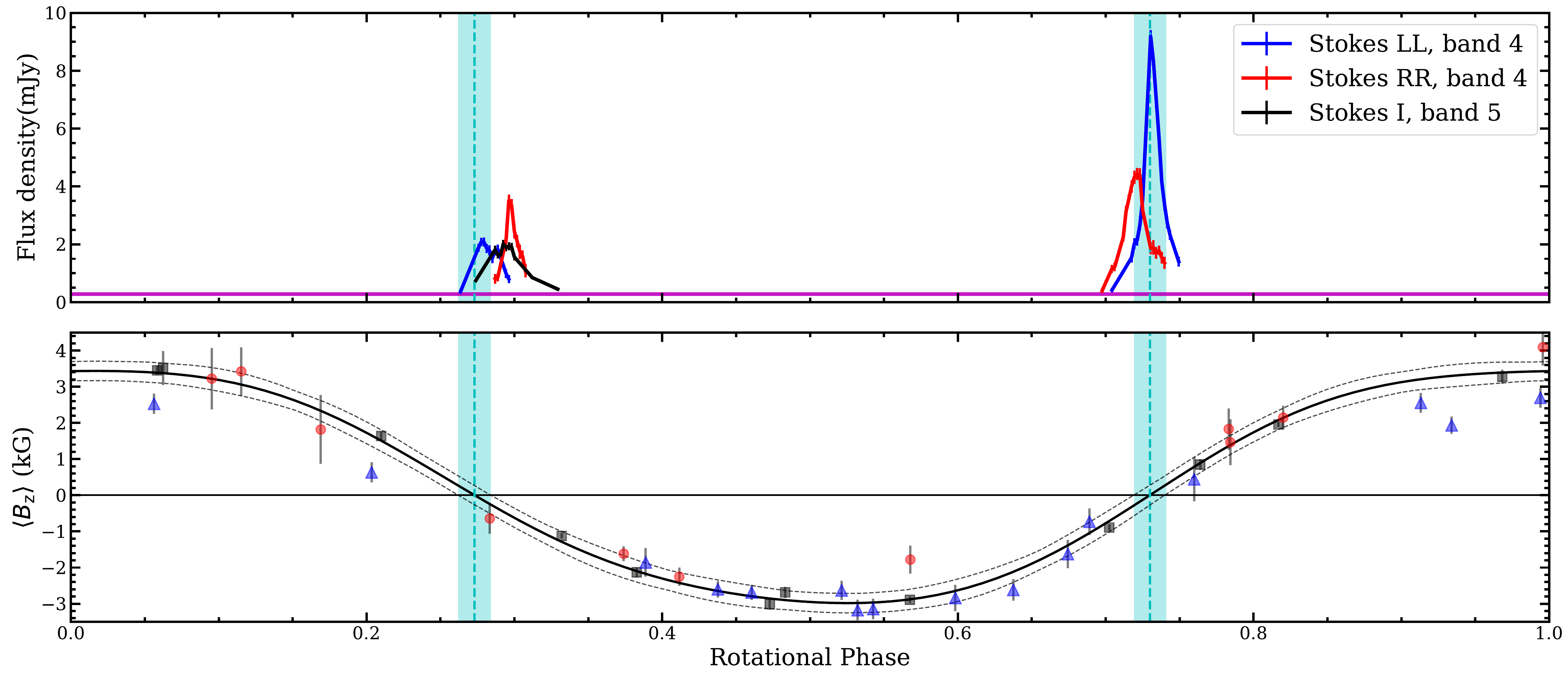}
    \caption{\textit{Upper panel:} Lightcurves of HD\,35298 phased with the ephemeris of \citet{shultz18}. The horizontal magenta line represents the approximate basal flux density for this star (see \S \ref{sec:hd35298} for details). This value is taken from \citet{linsky92} who reported the gyrosynchrotron flux density from this star (0.28$\pm$0.06 mJy) at 6 cm. \textit{Bottom panel:} The \bz~curve for the star. The \bz~data marked with black squares and red circles are taken from \citet{shultz18}; squares and circles correspond to data taken with the ESPaDOnS and dimaPol spectropolarimeters respectively. Note that the \bz~fit was performed using only the ESPaDOnS data. The data marked with blue triangles were taken from \citet{yakunin13}. The vertical dashed lines and the shaded regions represent the magnetic null phases and their uncertainties respectively.}
    \label{fig:hd35298_bz}
\end{figure*}

\section{Results}\label{sec:results}
The radio flux densities for HD\,35298 were phased using the following ephemeris \citep{shultz18}:
\begin{equation}
    \mathrm{HJD}=2454486.91(7)+1.85458(3)\cdot E
\end{equation}

We detected the star on all three days of observations (Figure \ref{fig:hd35298}). 
The data in band 5 cover the null at rotational phase 0.273. On all three days, we observed significant flux density enhancement. In the following two subsections, we describe the characteristics of the lightcurves observed in band 4 and band 5.

\subsection{Lightcurves in band 4}\label{subsec:res_band4}
In the case of the band 4 data, we see enhancements in both of the circular polarizations (Figure \ref{fig:hd35298}). The maximum observed circular polarization is $\approx$ 70\%. The pulse arrival sequences are opposite at the two nulls: near the magnetic null phase 0.273, where \bz~changes from positive to negative, the left circularly polarized (LCP) pulse arrives before the right circularly polarized (RCP) pulse, and near the other magnetic null phase (0.730), where \bz~changes from negative to positive, the RCP pulse arrives ahead of its LCP counterpart (Figure \ref{fig:hd35298_bz}). We attribute these enhancements to ECME because they are highly circularly polarized, occur near the magnetic nulls, and the pulse-arrival sequence for the RCP and LCP pulses is opposite at the two nulls, all as expected for ECME \citep{leto16}. Unlike the ideal ECME pulses depicted in Figure 6 of \cite{leto16}, the observed pulses are not symmetrically situated about the nulls, especially around the magnetic null phase 0.273. The mid point of the ECME pulses near this phase is offset by $\approx$ 0.016 cycle from the magnetic null. However the uncertainty in the null phase is 0.011 cycles and within this uncertainty, the magnetic null phase lies between the RCP and LCP pulses.

From the arrival sequence of the oppositely circularly polarized pulses near the two null phases, we deduce that the magneto-ionic mode of ECME is ordinary (O-mode) over the range of our observing frequency. The signatures of O-mode and X-mode ECME are illustrated in Figure 4 of \cite{leto19} and Figure 1 of \cite{das19}. Thus for our effective observing frequency range ($565-726$ MHz, once we exclude the edges of the band) $\nu_p/\nu_\mathrm{B}>0.3-0.35$, where $\nu_p$ and $\nu_\mathrm{B}$ are respectively the plasma frequency and the gyro-frequency at the region of emission \citep{melrose84,sharma84,leto19} assuming the emission to be in the fundamental mode. 
This dependence of the dominant mode of ECME on the ratio $\nu_\mathrm{p}/\nu_\mathrm{B}$ arises due to the fact that the cut-off frequencies for the two modes are functions of $\nu_\mathrm{p}$. Even though ECME favours the X-mode \citep{treumann06}, beyond a certain value of the plasma frequency, the X-mode cut-off becomes high enough to suppress the instability and the O-mode takes over \citep{sharma82}. The associated value of the ratio $\nu_\mathrm{p}/\nu_\mathrm{B}$ is around 0.3--0.35 near the transition region for a mildly relativistic electron distribution \citep{sharma84,melrose84}.
From this condition, we infer that $\nu_\mathrm{p}>194$ MHz (using $\nu_\mathrm{B}=645$ MHz) or $n_\mathrm{e}>5\times 10^8$ cm$^{-3}$ (using $\frac{\nu_\mathrm{p}}{\mathrm{Hz}}=9000\sqrt{\frac{n_\mathrm{e}}{\mathrm{cm^{-3}}}}$, where $n_\mathrm{e}$ is the electron number density) at the region of radio emission. 

\subsection{Lightcurve in band 5}\label{subsec:res_band5}
\begin{figure}
    \centering
    \includegraphics*[width=0.45\textwidth]{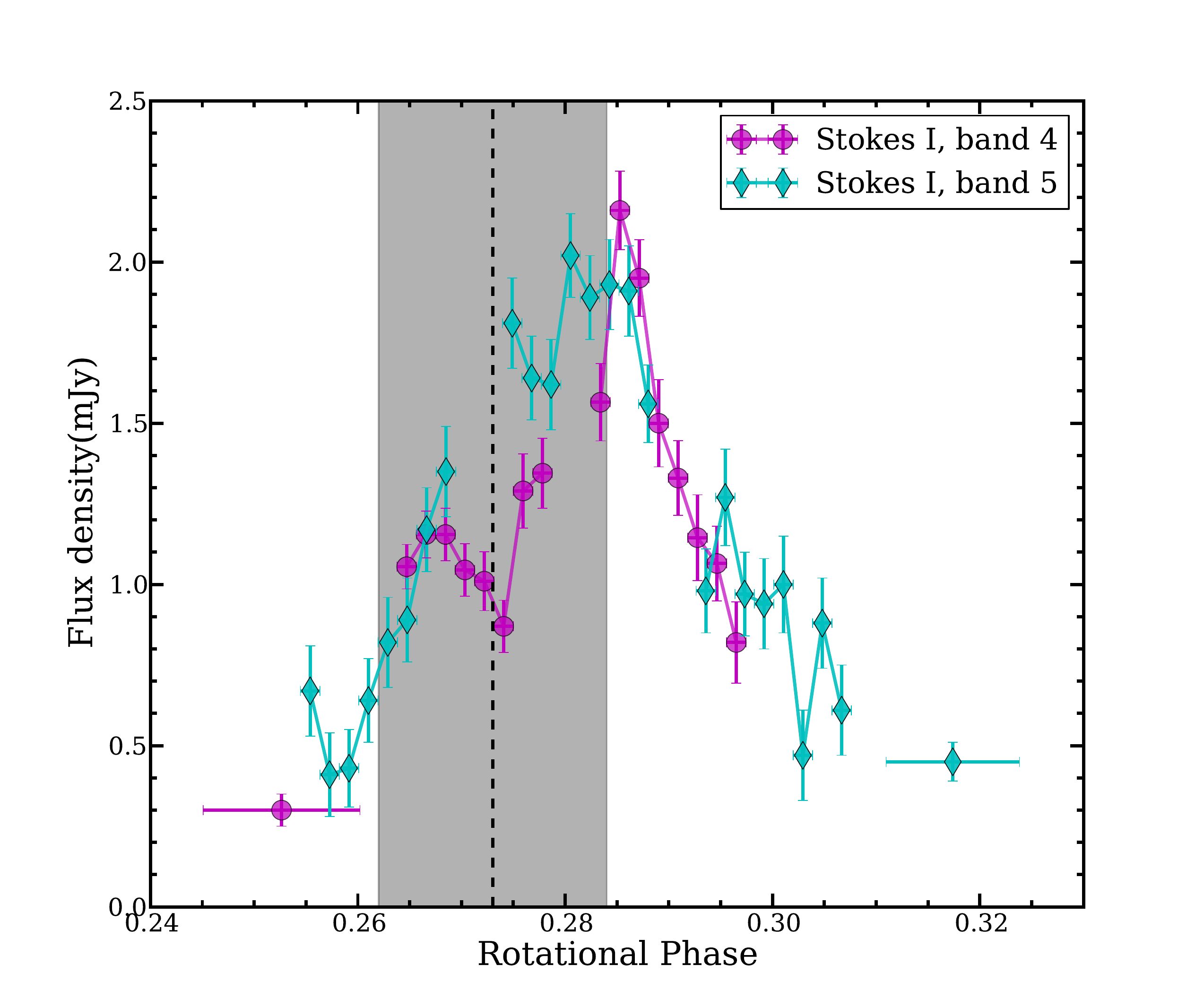}
    \caption{The total intensity (Stokes I) lightcurves of HD\,35298 near the magnetic null phase 0.273 (marked with a vertical dashed line) in band 4 and band 5. The shaded region around the vertical dashed line represents the uncertainty associated with the magnetic null phase.}
    \label{fig:hd35298_stokeI}
\end{figure}

Our observations in band 5 were inspired from our analysis of the L-band legacy GMRT data. The star was not detected in the legacy GMRT observation in 2014 which cover the magnetic null phase 0.273. HD\,35298 was one of the target sources and because of that there were large gaps between two consecutive scans of the star ($\approx 1$ hour). The scan which was closest to the magnetic null phase covered the phase window 0.282--0.289. The total phase coverage (ignoring the large gaps in between) was 0.137--0.319. We detected the star neither in the full time-averaged image, nor in the scan closest to the magnetic null. The 4$\sigma$ upper limit to the flux density of the star when its rotational phase was closest to the magnetic null phase is 1.6 mJy. This non-detection raises the question of whether the star has ECME at these frequencies or not. The absence of ECME at 1420 MHz would be a significant discovery since in all other stars (excluding HD\,133880), from which ECME has been observed, 1420 MHz lies inside the ECME bandwidth; in the case of HD\,133880, a strong enhancement in the L-band flux density, characteristic of ECME, was reported by \cite{chandra15}. 

In order to determine whether the non-detection in the archival L-band data is a result of insufficient sensitivity of the legacy GMRT, we observed the star in band 5 of the uGMRT (which is three times more sensitive than the legacy GMRT in band 5). 
In the new observation, we could detect the star throughout and there was a significant enhancement in its flux density. Comparing the band 5 lightcurve with the upper limit obtained from the archival L-band data (Figure \ref{fig:hd35298}, left panel), we found that the 4$\sigma$ upper limit is only slightly lower than the flux density observed in band 5. Since the strengths of the ECME pulses are known to vary between rotational cycles \citep[e.g.,][]{trigilio11}, we conclude that inadequate sensitivity of the legacy GMRT is a plausible reason for the non-detection of the star in the archival L-band data. The extreme case of the phenomenon of pulse strength variability was reported for CU\,Vir, where one of the ECME pulses at 13 cm was not detected at a certain epoch \citep{trigilio08}, then it was detected at a later epoch \citep{ravi10}, and disappeared once again when observed at a later epoch \citep{lo12}.


In Figure \ref{fig:hd35298_stokeI}, we show the total intensity (Stokes I) lightcurve of band 5 on top of the total intensity lightcurve in band 4. It can be clearly seen that the enhancements happen over the same range of rotational phases in both frequency bands.

\section{Discussion and Conclusion}\label{sec:disc_conc}
In this Letter, we report the discovery of the fifth main-sequence Bp type star (HD\,35298) to display ECME. Our observing frequency covers $565-726$ MHz (band 4 after removing the edges) and $1.13-1.38$ GHz (band 5 after removing the edges). Thus we confirm the presence of ECME at least over the frequency range of $0.56-1.38$ GHz. In future, we plan to observe this star in the S-band ($2-4$ GHz) of the Karl G. Jansky Very Large Array and band 3 ($260-500$ MHz) of the uGMRT to search for the upper and lower cut-offs of ECME respectively.

From the arrival sequence of the RCP and LCP pulses in band 4, we deduce that the mode of emission is the O-mode. This translates to a lower limit to the number density at the region of emission: $n_\mathrm{e}>5\times 10^8$ cm$^{-3}$ assuming emission at the fundamental harmonic. The corresponding height from the surface is $\approx 2.6$ $R_*\approx 6.3$ $R_\odot$. Previously, \cite{leto19} estimated the plasma density for HD\,142301 at a height where 1.5 GHz emission takes place, to be $10^9-10^{10}$ cm$^{-3}$. This corresponds to a height of 2 $R_*\approx 5$ $R_\odot$ from the stellar surface for the fundamental harmonic \citep[the stellar radius is $2.52$ $R_\odot$,][]{kochukhov06}. \cite{das19} suggested that the plasma density for HD\,142990 at a height of 1.4 $R_*\approx 4.1$ $R_\odot$ is likely to be $\sim 10^8$ cm$^{-3}$ (stellar radius is 2.8 $R_\odot$, Shultz et al., 2019 submitted). These estimates indicate that the magnetosphere of HD\,35298 is denser than that of HD\,142990, but probably similar (in terms of plasma density) to that of HD\,142301. Note that because of the assumption of emission at the fundamental harmonic, the estimated heights of ECME should only be regarded as the lower limits to their actual values.

With the discovery of ECME from HD\,35298, the number of hot magnetic stars known to show ECME now stands at five. Except for the very first star (CU\,Vir), the remaining four stars were discovered over a relatively short period of time ($\approx 1$ year). This naturally brings up the possibility of ECME being a common phenomenon among the magnetic A/B stars for a certain range of physical parameters. The physical quantity which is certain to play an important role in the presence or absence of ECME is the magnetic field. ECME has been observed from stars with surface magnetic fields ranging from 4 kG \citep[CU\,Vir,][]{kochukhov14} to 14 kG \citep[HD\,142301,][]{leto19}. The topology of the magnetic field has also been suggested to play a crucial role. In the case of $\sigma$ Ori E, the absence of ECME was attributed to the presence of a strong quadrupolar component \citep{leto12}. Interestingly, none of the 5 magnetic stars with ECME has been found to have an ideal dipolar magnetic field \citep[e.g.,][]{kochukhov14, kochukhov17, shultz18,leto19,glagolevskij10}. 
Thus, to solve the problem of when ECME can be stimulated, it will be important to have a precise constraints on the surface magnetic topology via Zeeman Doppler Imaging \citep[e.g.][]{piskunov02}.
Besides magnetic field, effective temperature ($T_\mathrm{eff}$), which correlates with the stellar wind, is likely to be an important parameter for ECME. The range of $T_\mathrm{eff}$ over which ECME has been observed is 12 kK (CU\,Vir, HD\,133880) to 18 kK (HD\,142990). 

Another physical quantity for which the role is not immediately evident is the stellar rotation period. All five ECME stars are relatively rapid rotators. HD\,35298 is the slowest ($P_\mathrm{rot}\approx 1.85$ days) among the five; the fastest is CU\,Vir \citep[$P_\mathrm{rot}\approx0.5$ days,][]{trigilio11}. 
However this could simply be a consequence of observational convenience. A rapid rotator requires less time to observe than a slower rotator. Moreover, for 3 of the 5 stars (CU\,Vir, HD\,142301 and HD\,142990), the pulses are found to be offset from their expected phases. Thus observation over a wider phase window is helpful in detecting ECME, which is easier to achieve in the case of a rapid rotator.

Frequency of observation is also important in detecting ECME from a star. Even if a star produces ECME, it will be present only over a certain range of frequencies. Until now, the highest frequency at which ECME from an early-type magnetic star has been observed is 2.5 GHz \citep[CU\,Vir,][]{trigilio08}, and the lowest frequency is 200 MHz \citep[HD\,142990,][]{lenc18}.  Looking at the fact that all the discoveries have been made at relatively low radio frequencies, we suggest that more observations of early-type magnetic stars below 2 GHz are likely to increase the number of hot magnetic stars known to exhibit ECME. 

In addition to these physical quantities, the distance of the star can also affect the ability to detect ECME. HD\,35298 is the most distant main-sequence star \citep[distance $\approx$ 371 pc, obtained from Gaia parallax,][]{gaiadr2_18} from which ECME has been observed. The second farthest star, HD\,142301, is at a distance of $\approx$ 148 pc (obtained from Gaia parallax). The corresponding radio luminosity (HD\,35298) lies in the range of $10^{17-18}$ erg s$^{-1}$ Hz$^{-1}$. For the cases of HD\,133880 and HD\,142990 also, the maximum observed radio luminosities lie in the same range. If we assume that $10^{18}$ erg s$^{-1}$ Hz$^{-1}$ is the typical maximum radio luminosity of ECME, we find that at a distance of around 1 kpc, the maximum ECME flux density will be reduced to $\sim$1 mJy over the frequency range of band 4. Thus beyond 1 kpc, it will be difficult to detect a complete ECME pulse.

To summarise, the discovery of ECME from 4 new stars in the last year, though encouraging, is not sufficient to tell us which subset of magnetic A/B stars is capable of producing ECME. More observations of these stars around their magnetic nulls will be needed to solve this problem.

\section*{Acknowledgements}
PC  acknowledges support from the Department of Science and Technology via 
SwarnaJayanti Fellowship awards (DST/SJF/PSA-01/2014-15). GAW acknowledges Discovery Grant support from the Natural Sciences and Engineering Research Council (NSERC) of Canada. MES acknowledges support from the Annie Jump Cannon Fellowship, supported by the University of Delaware and endowed by the Mount Cuba Astronomical Observatory. We thank the staff of the GMRT that made these observations possible. 
The GMRT is run by the National Centre for Radio Astrophysics of the Tata Institute 
of Fundamental Research. This research has made use of NASA's Astrophysics Data System.





\begin{thebibliography}{99}
\bibitem[Andre et al.(1988)]{andre88} Andre, P., Montmerle, T., Feigelson, E.~D., Stine, P.~C., \& Klein, K.-L.\ 1988, \apj, 335, 940




\bibitem[Chandra et al.(2015)]{chandra15} Chandra, P., Wade, G.~A., Sundqvist, J.~O., et al.\ 2015, \mnras, 452, 1245
\bibitem[Ciatti \& Bernacca(1971)]{ciatti71} Ciatti, F., \& Bernacca, P.~L.\ 1971, \aap, 11, 485 

\bibitem[Das et al.(2018)]{das18} Das, B., Chandra, P., \& Wade, G.~A.\ 2018, \mnras, 474, L61 

\bibitem[Das et al.(2019)]{das19} Das, B., Chandra, P., Shultz, M.~E., et al.\ 2019, \apj, 877, 123


\bibitem[Drake et al.(1987)]{drake87} Drake, S.~A., Abbott, D.~C., Bastian, T.~S., et al.\ 1987, \apj, 322, 902




\bibitem[Gaia Collaboration et al.(2018)]{gaiadr2_18} Gaia Collaboration, Brown, A.~G.~A., Vallenari, A., et al.\ 2018, \aap, 616, A1


\bibitem[Glagolevskij(2010)]{glagolevskij10} Glagolevskij, Y.~V.\ 2010, Astrophysics, 53, 536



\bibitem[Hallinan et al.(2006)]{hallinan06} Hallinan, G., Antonova, A., Doyle, J.~G., et al.\ 2006, \apj, 653, 690

\bibitem[Hallinan et al.(2007)]{hallinan07} Hallinan, G., Bourke, S., Lane, C., et al.\ 2007, \apjl, 663, L25

\bibitem[Hallinan et al.(2008)]{hallinan08} Hallinan, G., Antonova, A., Doyle, J.~G., et al.\ 2008, \apj, 684, 644

\bibitem[Hallinan et al.(2015)]{hallinan15} Hallinan, G., Littlefair, S.~P., Cotter, G., et al.\ 2015, \nat, 523, 568



\bibitem[Kao et al.(2016)]{kao16} Kao, M.~M., Hallinan, G., Pineda, J.~S., et al.\ 2016, \apj, 818, 24

\bibitem[Kochukhov \& Bagnulo(2006)]{kochukhov06} Kochukhov, O., \& Bagnulo, S.\ 2006, \aap, 450, 763 


\bibitem[Kochukhov et al.(2014)]{kochukhov14} Kochukhov, O., L{\"u}ftinger, T., Neiner, C., Alecian, E., \& MiMeS Collaboration 2014, \aap, 565, A83 


\bibitem[Kochukhov et al.(2017)]{kochukhov17} Kochukhov, O., Silvester, J., Bailey, J.~D., et al.\ 2017, \aap, 605, A13


\bibitem[Landstreet et al.(2007)]{landstreet07} Landstreet, J.~D., Bagnulo, S., Andretta, V., et al.\ 2007, \aap, 470, 685

\bibitem[Lenc et al.(2018)]{lenc18} Lenc, E., Murphy, T., Lynch, C.~R., Kaplan, D.~L., \& Zhang, S.~N.\ 2018, \mnras, 478, 2835 


\bibitem[Leone et al.(2004)]{leone04} Leone, F., Trigilio, C., Neri, R., et al.\ 2004, \aap, 423, 1095

\bibitem[Leto et al.(2006)]{leto06} Leto, P., Trigilio, C., Buemi, C.~S., Umana, G., \& Leone, F.\ 2006, \aap, 458, 831

\bibitem[Leto et al.(2012)]{leto12} Leto, P., Trigilio, C., Buemi, C.~S., Leone, F., \& Umana, G.\ 2012, \mnras, 423, 1766 

\bibitem[Leto et al.(2016)] {leto16} Leto, P., Trigilio, C., Buemi, C.~S., et al.\ 2016, \mnras, 459, 1159

\bibitem[Leto et al.(2017)]{leto17} Leto, P., Trigilio, C., Oskinova, L., et al.\ 2017, \mnras, 467, 2820

\bibitem[Leto et al.(2018)]{leto18} Leto, P., Trigilio, C., Oskinova, L.~M., et al.\ 2018, \mnras, 476, 562

\bibitem[Leto et al.(2019)]{leto19} Leto, P., Trigilio, C., Oskinova, L.~M., et al.\ 2019, \mnras, 482, L4


\bibitem[Linsky et al.(1992)]{linsky92} Linsky, J.~L., Drake, S.~A., \& Bastian T.~S.\ 1992, \apj, 393, 341

\bibitem[Lo et al.(2012)]{lo12} Lo, K.~K., Bray, J.~D., Hobbs, G., et al.\ 2012, \mnras, 421, 3316 


\bibitem[McMullin et al.(2007)]{mcmullin07} McMullin, J.~P., Waters, B., Schiebel, D., Young, W., \& Golap, K.\ 2007, Astronomical Data Analysis Software and Systems XVI, 376, 127

\bibitem[Melrose \& Dulk(1982)]{melrose82} Melrose, D.~B., \& Dulk, G.~A.\ 1982, \apj, 259, 844

\bibitem[Melrose et al.(1984)]{melrose84} Melrose, D.~B., Hewitt, R.~G., \& Dulk, G.~A.\ 1984, \jgr, 89, 897 



\bibitem[Mohanty \& Basri(2003)]{mohanty03} Mohanty, S., \& Basri, G.\ 2003, \apj, 583, 451




\bibitem[Petit et al.(2013)]{petit13} Petit, V., Owocki, S.~P., Wade, G.~A., et al.\ 2013, \mnras, 429, 398 

\bibitem[Piskunov, \& Kochukhov(2002)]{piskunov02} Piskunov, N., \& Kochukhov, O.\ 2002, \aap, 381, 736


\bibitem[Ravi et al.(2010)]{ravi10} Ravi, V., Hobbs, G., Wickramasinghe, D., et al.\ 2010, \mnras, 408, L99

\bibitem[Stelzer et al.(2006)]{stelzer06} Stelzer, B., Micela, G., Flaccomio, E., et al.\ 2006, \aap, 448, 293

\bibitem[Sharma et al.(1982)]{sharma82} Sharma, R.~R., Vlahos, L., \& Papadopoulos, K.\ 1982, \aap, 112, 377

\bibitem[Sharma \& Vlahos(1984)]{sharma84} Sharma, R.~R., \& Vlahos, L.\ 1984, \apj, 280, 405 

\bibitem[Shultz et al.(2018)]{shultz18} Shultz, M.~E., Wade, G.~A., Rivinius, T., et al.\ 2018, \mnras, 475, 5144 


\bibitem[Shultz et al.(2019)]{shultz19} Shultz, M.~E., Wade, G.~A., Rivinius, T., et al.\ 2019, \mnras, 485, 1508





\bibitem[Treumann(2006)]{treumann06} Treumann, R.~A.\ 2006, \aapr, 13, 229

\bibitem[Trigilio et al.(2000)]{trigilio00} Trigilio, C., Leto, P., Leone, F., Umana, G., \& Buemi, C.\ 2000, \aap, 362, 281

\bibitem[Trigilio et al.(2004)]{trigilio04} Trigilio, C., Leto, P., Umana, G., Leone, F., \& Buemi, C.S.\ 2004, \aap, 418, 593

\bibitem[Trigilio et al.(2008)]{trigilio08} Trigilio, C., Leto, P., Umana, G., Buemi, C.S., \& Leone, F.\ 2008, \mnras, 384, 1437

\bibitem[Trigilio et al.(2011)]{trigilio11} Trigilio, C., Leto, P., Umana, G., Buemi, C.S., \& Leone, F.\ 2011, \apjl, 739, L10





\bibitem[Yakunin(2013)]{yakunin13} Yakunin, I.~A.\ 2013, Astrophysical Bulletin, 68, 214
\end{thebibliography}





\bsp	

\label{lastpage}
\end{document}